\documentclass[conference, a4paper]{article}
\let\oldyear\year

\let\year\oldyear

\usepackage{cite}
\usepackage{amsmath,amssymb,amsfonts, amsthm}
\usepackage{algorithmic}
\usepackage{graphicx}
\usepackage{subfigure}
\usepackage{textcomp}
\def\BibTeX{{\rm B\kern-.05em{\sc i\kern-.025em b}\kern-.08em
    T\kern-.1667em\lower.7ex\hbox{E}\kern-.125emX}}

\usepackage[margin=0.9in]{geometry}
\usepackage[utf8]{inputenc}
\usepackage{complexity}
\usepackage{lineno}
\usepackage{authblk}
\usepackage{color}
\usepackage{cleveref}

\newcommand{\set}[1]{\{#1\}}
\newcommand{\inset}[2]{\{#1 \mid #2\}}

\newtheorem{theorem}{Theorem}
\newtheorem{problem}{Problem}

\newcommand{\ourproblem}{\textsc{MinBraiding}}
\newcommand{\rlsat}{\textsc{PlanarRectilinear3SAT}}
\newcommand{\psat}{\textsc{Planar3SAT}}
\newcommand{\sat}{\textsc{3SAT}}
\newcommand{\lit}[2]{\ell_{#1}^{#2}}
\newcommand{\Yes}{\texttt{Yes}}

\newcommand{\vgate}[2]{X_{#1}^{\texttt{v}}(#2)}
\newcommand{\vgateset}[1]{\mathcal{R}^{\texttt{v}}(#1)}

\newcommand{\copygate}[2]{X_{#1}^\texttt{p}(#2)}
\newcommand{\bendgate}[2]{X_{#1}^\texttt{b}(#2)}
\newcommand{\fixgate}[2]{X_{#1}^\texttt{f}(#2)}

\usepackage{tikz}
\usepackage{ifthen}
\def\myscale{0.6}
\usetikzlibrary{positioning}
\usetikzlibrary{calc}

\tikzstyle{fixed}=[line width=10*\myscale, color=gray]
\tikzstyle{variable bar}=[line width=10*\myscale, color=teal]
\tikzstyle{inner bar}=[line width=10*\myscale, color=orange]
\tikzstyle{vertical bar}=[line width=2*\myscale, color=lightgray]
\makeatletter
\tikzset{
    from/.style args={#1 to #2}{
        above right={0cm of #1},
        /utils/exec=\pgfpointdiff
            {\tikz@scan@one@point\pgfutil@firstofone(#1)\relax}
            {\tikz@scan@one@point\pgfutil@firstofone(#2)\relax},
        minimum width/.expanded=\the\pgf@x,
        minimum height/.expanded=\the\pgf@y}}
\makeatother

\newcommand{\vgadget}[4]{
    \coordinate (blv#4) at (0 + #1, 0 + #2 - 1);
    \coordinate (urv#4) at (3 + #1, 3 + #2 + 1);
    \draw[draw = none, fill=blue!40, rounded corners=3pt] (blv#4) rectangle (urv#4);

    \foreach \x in { 0.5, ..., 2.5} {
            \draw [vertical bar] (\x + #1, -1+#2) to (\x + #1, 4 + #2);
        }
    \foreach \l / \r / \y in {
            0/3/3,
            0/1/2,
            0/1/1,
            0/3/0
        } {
            \draw [fixed] (\l-0.2 + #1, \y + #2) to (\r+0.2 + #1, \y + #2);
        }
    \foreach \l / \r / \y in {
            2/3/2
        } {
            \draw [variable bar] (\l-0.2 + #1, \y + #2 - 1 + #3) to (\r+0.2 + #1, \y + #2 - 1 + #3);
        }
    \draw[draw = black, dotted, rounded corners=3pt] (blv#4) rectangle (urv#4);
}

\newcommand{\drawgate}[7]{
    \ifnum#6>0
        \draw [#1] (#2-0.2, #3) to (#4+0.2, #5);
    \else
        \draw [#1] (#7-#2 + 0.2, #3) to (#7-#4 - 0.2, #5);
    \fi
}

\newcommand{\cgadget}[9]{
    \coordinate (blc#9) at (0 + #1 - 1, 0 + #2 - 1);
    \coordinate (urc#9) at (8 + #1,     10 + #6 + #7 + #2 + 1);
    \draw[draw=none, fill=yellow!40, rounded corners=3pt] (blc#9) rectangle (urc#9);
    \foreach \x in {-0.5, 0.5, ..., 7.5} {
            \draw [vertical bar] (\x + #1, -1 + #2) to (\x + #1, 11 + #6 + #7 + #2);
        }

    \foreach \l / \r / \y in {
            -1/8/10 + #6 + #7,
            7/8/9   + #6 + #7,
            7/8/8   + #6 + #7,
            -1/4/7 + #7, 7/8/7 + #7,
            7/8/6 + #7,
            7/8/5 + #7,
            -1/1/4, 7/8/4,
            -1/3/3, 7/8/3,
            7/8/2,
            7/8/1,
            -1/8/0} {
            \drawgate{fixed}{\l + #1}{\y + #2}{\r + #1}{\y + #2}{#8}{7}
        }

    \foreach \l / \r / \y in {
            -1/0/9 - 1 + #3 + #6 + #7,
            -1/0/6 - 1 + #4 + #7,
            -1/0/2 - 1 + #5
        } {
            \drawgate{variable bar}{\l + #1}{\y + #2}{\r + #1}{\y + #2}{#8}{7}
        }

    \foreach \l / \r / \y in {
            1/6/9 - 1 + #3 + #6 + #7,
            5/6/8 - 1 + #3 + #6 + #7,
            5/6/7 - 1 + #3 + #6 + #7,
            5/6/6 - 1 + #3 + #6 + #7,
            0/1/8 + 1 - #3 + #6 + #7,
            1/3/6 - 1 + #4 + #7,
            2/4/5 - 1 + #4 + #7,
            0/1/5 + 1 - #4 + #7,
            4/6/3 + 1 - #5 + 1 - #4 - #5 + #4 * #5 + #7 - #4 * #7 - #5 * #7 + #4 * #5 * #7,
            5/6/2 + 1 - #5,
            4/6/1 + 1 - #5,
            1/4/2 - 1 + #5,
            0/1/1 + 1 - #5
        } {
            \drawgate{inner bar}{\l + #1}{\y + #2}{\r + #1}{\y + #2}{#8}{7}
        }

    \ifnum#7>0
        \foreach \ey in {1, ..., #7}{
                \foreach \l / \r / \y in {
                        -1/1/4 + \ey, 7/8/4 + \ey
                    } {
                        \drawgate{fixed}{\l + #1}{\y + #2}{\r + #1}{\y + #2}{#8}{7}
                    }
                \foreach \l / \r / \y in {
                        2/4/5 - 1 + #4 + #7 - \ey
                    } {
                        \drawgate{inner bar}{\l + #1}{\y + #2}{\r + #1}{\y + #2}{#8}{7}
                    }
            }
    \else
    \fi

    \ifnum#6>0
        \foreach \ey in {1, ..., #6}{
                \foreach \l / \r / \y in {
                        -1/4/7 + #7 + \ey, 7/8/7 + #7 + \ey
                    } {
                        \drawgate{fixed}{\l + #1}{\y + #2}{\r + #1}{\y + #2}{#8}{7}
                    }
                \foreach \l / \r / \y in {
                        5/6/6 - 2 + #3 + \ey + #7
                    } {
                        \drawgate{inner bar}{\l + #1}{\y + #2}{\r + #1}{\y + #2}{#8}{7}
                    }
            }
    \else
    \fi
    \draw[draw=black, dashed,  rounded corners=3pt] (blc#9) rectangle (urc#9);
}

\newcommand{\copygadget}[5]{
    \foreach \x in {1.5, 2.5, ..., 9.5} {
            \draw [vertical bar] (\x + #1, -1 + #2) to (\x + #1, 10 + #2 + #4 * 2);
        }
    \foreach \l / \r / \y in {
            1/10/9 + #4 * 2,
            1/2/8 + #4 * 2,
            8/10/8 + #4 * 2,
            1/2/7 + #4 * 2,
            8/10/5 + #4 * 2,
            1/2/4, 8/10/4,
            1/2/3,
            1/2/2,
            1/2/1, 8/10/1,
            1/10/0} {
            \drawgate{fixed}{\l + #1}{\y + #2}{\r + #1}{\y + #2}{#5}{11}
        }
    \foreach \l / \r / \y in {
            9/10/7 - 1 + #3 + #4 * 2,
            9/10/3 - 1 + #3,
            1/2/6  - 1 + #3 + #4 * 2
        } {
            \drawgate{variable bar}{\l + #1}{\y + #2}{\r + #1}{\y + #2}{#5}{11}
        }

    \foreach \l / \r / \y in {
            3/5/7 + 1 - #3 + #4 * 2,
            3/4/6 + 1 - #3 + #4 * 2,
            8/9/6 + 1 - #3 + #4 * 2,
            2/5/5 + 1 - #3 + #4 * 2,
            3/4/4 + 1 - #3,
            3/5/3 + 1 - #3,
            3/4/2 + 1 - #3,
            8/9/2 + 1 - #3,
            3/5/1 + 1 - #3,
            5/7/8 - 1 + #3 + #4 * 2,
            6/8/7 - 1 + #3 + #4 * 2,
            5/7/6 - 1 + #3 + #4 * 2,
            6/7/5 - 1 + #3 + #4 * 2,
            5/7/4 - 1 + #3,
            6/8/3 - 1 + #3,
            5/7/2 - 1 + #3
        } {
            \drawgate{inner bar}{\l + #1}{\y + #2}{\r + #1}{\y + #2}{#5}{11}
        }
    \ifnum#4>0
        \foreach \ey in {1, ..., #4}{
                \foreach \l / \r / \y in {
                        1/2/4 + 2 * \ey, 8/10/4 + 2* \ey,
                        1/2/4 + 2 * \ey - 1, 8/10/4 + 2* \ey - 1
                    } {
                        \drawgate{fixed}{\l + #1}{\y + #2}{\r + #1}{\y + #2}{#5}{11}
                    }
                \foreach \l / \r / \y in {
                        3/4/4 + 1 - #3 + 2 * \ey, 5/7/4 - 1 + #3 + 2 * \ey,
                        3/5/4 + 1 - #3 + 2 * \ey - 1, 6/7/4 - 1 + #3 + 2 * \ey - 1
                    } {
                        \drawgate{inner bar}{\l + #1}{\y + #2}{\r + #1}{\y + #2}{#5}{11}
                    }
            }
    \else
    \fi
}

\newcommand{\bendgadget}[5]{
    \foreach \x in {-0.5, 0.5, ..., 8.5} {
            \draw [vertical bar] (\x + #1, -1 + #2 ) to (\x + #1, 8 + #2 + #4 * 2);
        }
    \foreach \l / \r / \y in {
            -1/9/7 + #4 * 2,
            8/9/6 + #4 * 2,
            8/9/5 + #4 * 2,
            -1/2/4,
            8/9/4,
            8/9/3,
            8/9/2,
            -1/2/1,
            8/9/1,
            -1/9/0
        } {
            \drawgate{fixed}{\l + #1}{\y + #2}{\r + #1}{\y + #2}{#5}{8}
        }

    \foreach \l / \r / \y in {
            -1/0/6 - 1 + #3 + #4 * 2,
            -1/0/3 - 1 + #3
        } {
            \drawgate{variable bar}{\l + #1}{\y + #2}{\r + #1}{\y + #2}{#5}{8}
        }

    \foreach \l / \r / \y in {
            1/2/6 - 1 + #3 + #4 * 2,
            5/7/6 - 1 + #3 + #4 * 2,
            6/7/5 - 1 + #3 + #4 * 2,
            5/7/4 - 1 + #3,
            1/2/3 - 1 + #3,
            6/7/3 - 1 + #3,
            5/7/2 - 1 + #3,
            0/1/5 + 1 - #3 + #4 * 2,
            2/5/5 + 1 - #3 + #4 * 2,
            3/4/4 + 1 - #3,
            3/5/3 + 1 - #3,
            0/1/2 + 1 - #3,
            2/4/2 + 1 - #3,
            3/5/1 + 1 - #3
        } {
            \drawgate{inner bar}{\l + #1}{\y + #2}{\r + #1}{\y + #2}{#5}{8}
        }
    \ifnum#4>0
        \foreach \ey in {1, ..., #4}{
                \foreach \l / \r / \y in {
                        -1/2/4 + \ey * 2,
                        8/9/4 + \ey * 2,
                        -1/2/4 + \ey * 2 - 1,
                        8/9/4 + \ey * 2 - 1
                    } {
                        \drawgate{fixed}{\l + #1}{\y + #2}{\r + #1}{\y + #2}{#5}{8}
                    }
                \foreach \l / \r / \y in {
                        3/4/4 + 1 - #3 + \ey * 2,
                        5/7/4 - 1 + #3 + \ey * 2,
                        3/5/4 + 1 - #3 + \ey * 2 - 1,
                        6/7/4 - 1 + #3 + \ey * 2 - 1
                    } {
                        \drawgate{inner bar}{\l + #1}{\y + #2}{\r + #1}{\y + #2}{#5}{8}
                    }
            }
    \else
    \fi
}

\newcommand{\notgadget}[3]{

    \foreach \x in {1.5, 2.5, ..., 4.5} {
            \draw [vertical bar] (\x + #1, -1 + #2) to (\x + #1, 4 + #2);
        }
    \foreach \l / \r / \y in {
            1/5/3,
            1/5/0
        } {
            \drawgate{fixed}{\l + #1}{\y + #2}{\r + #1}{\y + #2}{1}{0}
        }
    \foreach \l / \r / \y in {
            1/2/2 - 1 + #3,
            4/5/1 + 1 - #3
        } {
            \drawgate{variable bar}{\l + #1}{\y + #2}{\r + #1}{\y + #2}{1}{0}
        }

    \foreach \l / \r / \y in {
            3/4/2 - 1 + #3,
            2/3/1 + 1 - #3
        } {
            \drawgate{inner bar}{\l + #1}{\y + #2}{\r + #1}{\y + #2}{1}{0}
        }
}

\newcommand{\extendgadget}[4]{
    \foreach \x in {1, 2, ..., #4} {
            \draw [vertical bar] (\x + #1 + 0.5, -1 + #2) to (\x + #1 + 0.5, 4 + #2);
        }
    \foreach \l / \r / \y in {
            1/5 + #4 - 4/3,
            1/5 + #4 - 4/0
        } {
            \drawgate{fixed}{\l + #1}{\y + #2}{\r + #1}{\y + #2}{1}{0}
        }
    \foreach \l / \r / \y in {
            1/5+#4-4/2 - 1 + #3
        } {
            \drawgate{variable bar}{\l + #1}{\y + #2}{\r + #1}{\y + #2}{1}{0}
        }

}

\newcommand{\dummyvgadget}[3]{
    \foreach \x in { 0.5, ..., 2.5} {
            \draw [vertical bar] (\x + #1, -1+#2) to (\x + #1, 4 + #2);
        }
    \foreach \l / \r / \y in {
            0/3/3,
            0/1/2,
            0/1/1,
            0/3/0
        } {
            \drawgate{fixed}{\l + #1}{\y + #2}{\r + #1}{\y + #2}{#3}{3}
        }
    \foreach \l / \r / \y in {
            2/3/2,
            2/3/1
        } {
            \drawgate{variable bar}{\l + #1}{\y + #2}{\r + #1}{\y + #2}{#3}{3}
        }
}

\newcommand{\fixgadget}[3]{
    \foreach \x in {0, ..., #3} {
            \draw [vertical bar] (\x + #1 + 0.5, -1 + #2) to (\x + #1 + 0.5, 7 + #2);
        }
    \foreach \l / \r / \y in {
            0/#3 + 1/6,
            0/#3 - 1/5,
            #3/#3 + 1/5,
            0/#3 - 2/4,
            #3 - 1/#3 + 1/4,
            0/2/1,
            3/#3 + 1/1,
            0/1/0,
            2/#3 + 1/0
        } {
            \drawgate{fixed}{\l + #1}{\y + #2}{\r + #1}{\y + #2}{1}{0}
        }
    \node [circle, fill, inner sep=0, minimum width=2pt] at (#1 + #3/2 + 0.5, #2 + 2.5 - 0.5){};
    \node [circle, fill, inner sep=0, minimum width=2pt] at (#1 + #3/2 + 0.5, #2 + 2.5){};
    \node [circle, fill, inner sep=0, minimum width=2pt] at (#1 + #3/2 + 0.5, #2 + 2.5 + 0.5){};
}

\begin{document}

\title{Hardness of braided quantum circuit optimization\\ in the surface code}
\author[1]{Kunihiro Wasa \footnote{email: kunihiro.wasa@gmail.com}}
\author[2,3,4]{Shin Nishio}
\author[3]{Koki Suetsugu}
\author[5]{Michael Hanks}
\author[3]{Ashley~Stephens}
\author[3]{Yu~Yokoi}
\author[4,3,2]{Kae Nemoto}

\affil[1]{Faculty of Science and Engineering, Hosei University, Kajino-cho 3-7-2, Koganei-shi, Tokyo, 184-8584, Japan}
\affil[2]{Department of Informatics, School of Multidisciplinary Sciences, SOKENDAI (The Graduate University for Advanced Studies), 2-1-2 Hitotsubashi, Chiyoda-ku, Tokyo, 101-8430, Japan}
\affil[3]{National Institute of Informatics, 2-1-2 Hitotsubashi, Chiyoda-ku, Tokyo, 101-8430, Japan}
\affil[4]{Quantum Information Science and Technology Unit, Okinawa Institute of Science and Technology Graduate University, Onna-son, Kunigami-gun, Okinawa, 904-0495, Japan}
\affil[5]{QOLS, Blackett Laboratory, Imperial College London, London SW7 2AZ, United Kingdom}
\date{} 

\maketitle

\begin{abstract}
Large-scale quantum information processing requires the use of quantum error correcting codes to mitigate the effects of noise in quantum devices. Topological error-correcting codes, such as surface codes, are promising candidates as they can be implemented using only local interactions in a two-dimensional array of physical qubits. Procedures such as defect braiding and lattice surgery can then be used to realize a fault-tolerant universal set of gates on the logical space of such topological codes. However, error correction also introduces a significant overhead in computation time, the number of physical qubits, and the number of physical gates. While optimizing fault-tolerant circuits to minimize this overhead is critical, the computational complexity of such optimization problems remains unknown. This ambiguity leaves room for doubt surrounding the most effective methods for compiling fault-tolerant circuits for a large-scale quantum computer. In this paper, we show that the optimization of a special subset of braided quantum circuits is NP-hard by a polynomial-time reduction of the optimization problem into a specific problem called \rlsat.\\

\noindent Index terms: Braiding circuits, computational complexity, fault-tolerant quantum computation, NP-complete, planar rectilinear 3SAT, quantum circuit optimization, surface code
\end{abstract}

\section{Introduction}
Quantum information processing has many applications, including quantum computation\cite{divincenzo1995quantum,grover1996fast,shor1994algorithms} and quantum communication\cite{briegel1998quantum, bennett2020quantum}. Implementations of Noisy Intermediate Scale Quantum (NISQ) systems already exist, but large-scale quantum information processing is required for the majority of proposed applications. Noise in these systems is an obstacle to this, and quantum error correcting codes (QECCs) are the most promising methods to address this issue. QECCs provide a way of detecting and correcting errors in quantum systems by storing quantum information in a redundant system\cite{shor1995scheme}. The ability to perform reliable computation in the presence of errors is called fault-tolerant quantum computation (FTQC)\cite{gottesman1997stabilizer}. Topological stabilizer codes such as surface codes\cite{kitaev1997quantum} are good candidates for implementing FTQC due to their high thresholds and low-weight local stabilizers\cite{raussendorf2007fault, stephens2014fault}. 

However, many additional resources are required to implement fault-tolerant error correction, including in the case of surface codes due to their low encoding rate\cite{bravyi2010tradeoffs, jones2012layered, gidney2021factor}. Furthermore, fault-tolerant quantum circuits are longer than unencoded ones since it is only possible to implement a finite set of gates fault-tolerantly within a given code and circuit depth increases significantly when continuous quantum gates are decomposed into gates from such a set  \cite{dawson2005solovay}. For universal quantum computation, it is necessary to implement a set of quantum gates called a universal gate set\cite{deutsch1989quantum, barenco1995elementary, divincenzo1995two, barenco1995universal, lloyd1995almost,deutsch1995universality, reck1994experimental} which must contain at least one quantum gate for multiple qubits. Transversal implementations of these gates in 2D topological codes require non-local connectivity or higher-dimensional architectures, and so alternative techniques such as defect braiding\cite{fowler2012surface} and lattice surgery\cite{horsman2012surface} have been proposed. These operations can also take a long time to perform compared to performing logic on unencoded qubits and therefore optimization of these quantum circuits can significantly reduce the overhead for FTQC. 

This project deals with the problem of optimizing defect braiding circuits to achieve FTQC using surface codes. Several practical optimizations for braiding circuits have been proposed\cite{paetznick2013quantum, paler2017fault, hua2021autobraid, hanks2020effective}, but in order to achieve further optimizations it is helpful to classify the optimization problem and prove the computational complexity. This allows us to estimate the cost of the optimization problem and define the requirements for the computer architecture. Our results show that the computational complexity for optimizing the depth of the defect braiding circuits by reordering is \NP-hard, which matches previously known results for lattice surgery\cite{herr2017optimization}.

The surface code can be defined by placing qubits on the edges of a regular square lattice. Error detection is then achieved by the measurement of multi-qubit Pauli operators localized at the nodes and across the faces of this lattice, which define the encoded subspace. Logical qubits are typically associated with lattice defects such as boundaries or punctures, which correspond to regions of the lattice remaining unmeasured. In this work we consider an encoding where a logical qubit is encoded with a pair of such punctures. Logical operations between pairs of logical qubits can then be performed via braiding of these pairs. Some restrictions are imposed on the braiding operations that can be performed, both by the topological nature of the surface code and the requirement to maintain a sufficient degree of error correction. For the purpose of our problem, the relevant restrictions are that defects must be separated by a minimum distance and that braiding operations do not commute.

\begin{figure}[htbp]
  \begin{minipage}[b]{0.33\linewidth}
    \centering
    \includegraphics[keepaspectratio, width = 0.8\linewidth]{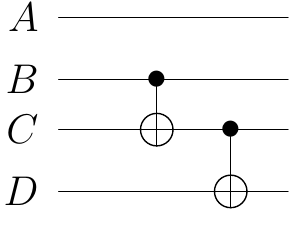}
    \label{simple_cx}
    \begin{center}
        (a)
    \end{center}
  \end{minipage}
  \begin{minipage}[b]{0.33\linewidth}
    \centering
    \includegraphics[keepaspectratio,width = 0.99\linewidth]{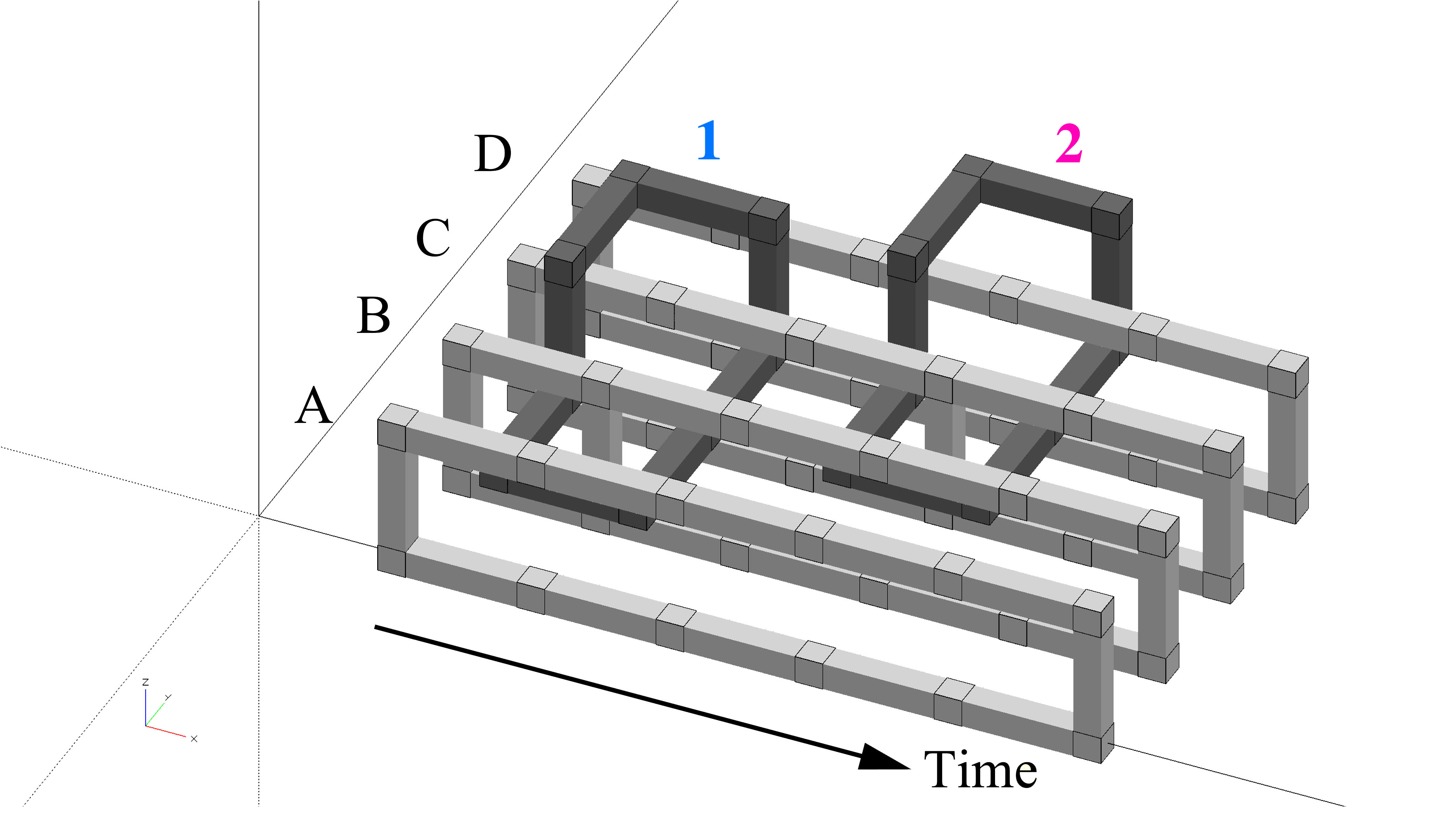}
    \begin{center}
        (b)
    \end{center}
    \label{3D_defect}
  \end{minipage}
  \begin{minipage}[b]{0.33\linewidth}
    \centering
    \begin{tikzpicture}[scale=\myscale]
	\foreach \x in {0.5, 1.5, 2.5, 3.5} {
		\draw [vertical bar] (\x, 0) to (\x, 4);
	}
	\node [black] at (0.3 + 0.3*\myscale, 4.5) {$A$}; 
	\node [black] at (1.3 + 0.3*\myscale, 4.5) {$B$}; 
	\node [black] at (2.3 + 0.3*\myscale, 4.5) {$C$}; 
	\node [black] at (3.3 + 0.3*\myscale, 4.5) {$D$}; 
	\foreach \l / \r / \y in {
        1/3/3
	} {
		\draw [variable bar, darkgray] (\l-0.2, \y) to (\r+0.2, \y); 
	}

    \node [blue] at (3.5 + 0.3*\myscale, 3) {$1$}; 
	\foreach \l / \r / \y in {
        2/4/2
	} {
		\draw [variable bar, darkgray] (\l-0.2, \y) to (\r+0.2, \y); 
	}

    \node [magenta] at (4.5 + 0.3*\myscale, 2) {$2$}; 

    \draw (-1, 4) edge[->] (-1, -0.5); 
\end{tikzpicture}
    \begin{center}
        (c)
    \end{center}
    \label{defect_diagram}
  \end{minipage}
  \caption{(a) A simple example of a quantum circuit with $\textrm{CX}$ gates. (b) Defect braiding operation that implements the circuit of (a). Time flows from left to right, and the gray bars labeled with letters A to D are the defect pairs corresponding to the logical qubits. The black loops $1$ and $2$ correspond to the braiding operations ($\textrm{CX}$ gates). (c) Simplified diagram (1D braiding circuit representation) of (b). We ignore the direction of the $\textrm{CX}$ gate for simplicity. Time flows from top to bottom. Vertical gray bars correspond to the logical qubits (pairs of punctures) and the black boxes correspond to the logical gate operations.}
  \label{defect_diagrams}
\end{figure}
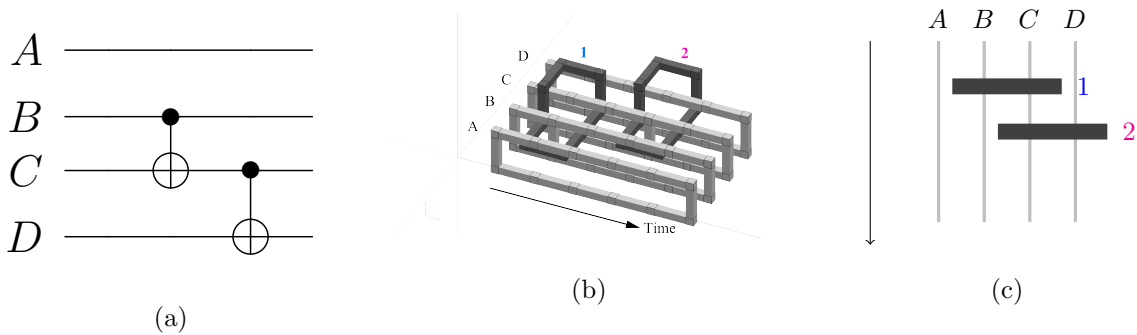

A simple quantum circuit is presented in Figure~\ref{defect_diagrams}(a) as an example.
Figure~\ref{defect_diagrams}(b) then shows a braided implementation as a 3D space-time diagram, with a 1D arrangement of logical qubits, while Figure~\ref{defect_diagrams}(c) shows a corresponding simplified diagram that we call the ``1D braiding circuit'' representation. 
More generally, braiding circuits can be implemented in a high-dimensional representational space, and by imposing the constraints we obtain the 1D braiding circuit. These braiding circuits are then implemented on a certain quantum computer architecture, where the quantum data qubits are often in one-dimensional arrangement. This is due to the large distillation factory associated with this code implementation\cite{devitt2013requirements}.

In this paper, we investigate the computational complexity of the problem of minimizing the depth of the 1D braiding circuit.
For this purpose, we formulate an optimization problem \ourproblem{}, which asks whether there exists an arrangement of qubits and gates that attains the required processing time. 

\clearpage
\section{Problem definition}
Let $[n]=\{1,2,\dots,n\}$ denote the set of $n$ logical qubits. 
We are given a family $\mathcal{R}$ of controlled-NOT gates, where each gate is represented by $X\in \mathcal{R}$, a subset of $[n]$ corresponding to the set of qubits to which the gate is applied. 
We are also given a partial order $\succeq$ on $\mathcal{R}$, which represents the order of operators on the same qubit.
Sometimes gate operators on the same qubit are commutative, but for simplicity we ignore such commutativity in this paper.
Therefore two gates $X, X'\in \mathcal{R}$ are comparable if and only if they share a qubit. 
That is, we have $X\succeq X'$ or $X'\succeq X$ if $X\cap X'\neq \emptyset$. 
We write $X\succ X'$ if $X\succeq X'$ and $X\neq X'$.
For a permutation $\pi:[n]\to [n]$ of logical qubits, we say that gates $X_1, \dots, X_\ell\in \mathcal{R}$ can be \emph{arranged} into one row with respect to $\pi$ if, for any distinct $i,j\in \{1,2,\dots,\ell\}$, 
either $\max\inset{\pi(x)}{x\in X_i}+1<\min\inset{\pi(x')}{x'\in X_j}$
or $\max\inset{\pi(x')}{x'\in X_j}+1<\min\inset{\pi(x)}{x\in X_i}$ holds.
This condition means that any pair of gates in the same row have no overlap, and furthermore there is a margin between them.

We say that the family $\mathcal{R}$ of gates can be {\em packed} with height $h\in \mathbf{Z}_{+}$ according to a partial order $\succeq$ if there exists a pair $(\pi,\mu)$ of a permutation $\pi:[n]\to [n]$ and a function $\mu:\mathcal{R} \to [h]$ satisfying the following two conditions:
\begin{itemize}
    \item[(i)] for each $i\in [h]$, the gates in $\inset{X}{\mu(X)=i}$ can be arranged into one row with respect to a permutation $\pi:[n]\to [n]$, and
    \item[(ii)] for any two distinct gates $X, X'\in \mathcal{R}$, if $X\cap X'\neq \emptyset$ and $X\succ X'$, 
    then $\mu(X)>\mu(X')$.
\end{itemize}
For $\pi$ and $\mu$ satisfying (i) and (ii), a pair $(\pi, \mu)$ is called a \emph{packing} of $(\mathcal{R},\succeq)$ with height $h$.
We call condition (i) the \emph{horizontal condition}, and condition (ii) the \emph{vertical condition}. 
For a packing $(\pi, \mu)$, we call $\mu(X)$ the \emph{level of $X\in \mathcal{R}$}. 

Now, we define our problem, which we call \ourproblem{}.
\begin{problem}[\ourproblem{}]
Given a set $\mathcal{R}$ of gates, a partial order $\succeq$ on $\mathcal{R}$, and a positive integer $h$, 
output \Yes{} if $\mathcal{R}$ can be packed according to $\succeq$ with height at most $h$. 
\end{problem}

In particular, if there is a packing $(\pi, \mu)$ of $(\mathcal{R},\succeq)$ with height $h$ but no packing with height $h' < h$, then we call $(\pi, \mu)$ a \emph{minimum packing}. 

\section{Hardness result}

In this section, we show that \ourproblem{} is \NP-complete. 
Clearly, \ourproblem{} is in \NP. 
Therefore, we devote this section to showing the polynomial-time reduction from \rlsat{}, which is explained below. 

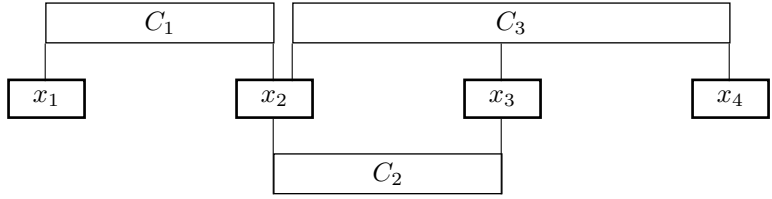
\begin{figure*}[t]
    \centering
    \begin{tikzpicture}
        \node[rectangle, draw=black, line width=1pt, from={(0,0) to (1,0.5)}]  {$x_1$};
        \node[rectangle, draw=black, line width=1pt, from={3,0 to 4,0.5}]  {$x_2$};
        \node[rectangle, draw=black, line width=1pt, from={6,0 to 7,0.5}]  {$x_3$};
        \node[rectangle, draw=black, line width=1pt, from={9,0 to 10,0.5}] {$x_4$};

        \foreach \x / \y / \w / \c in {
        0.5/1/3/$C_1$, 
        3.75/1/5.75/$C_3$, 
        3.5/-1/3/$C_2$} {
        \node[rectangle, draw=black, line width=0.5pt, from={(\x, \y) to (\x+\w, \y+0.5)}]  {\c};
        }
        
        \foreach \x  in {0.5, 3.5, 3.75, 6.5, 9.5} {
            \draw (\x, 1) -- (\x, 0.5); 
        }
        \foreach \x  in {3.5, 6.5} {
            \draw (\x, -1) -- (\x, 0); 
        }
\end{tikzpicture}
       \caption{Example of the rectilinear representation of an instance
    $\phi_1 =  C_1 \land C_2 \land C_3$ of \rlsat{}, where $C_1 =(\lnot x_1 \lor x_2), C_2 = (x_2 \lor x_3 ), C_3 = (x_2 \lor x_3 \lor \lnot x_4)$. }
    \label{fig:ex.rectilinear}
\end{figure*}

A 3-CNF Boolean formula $\phi$ consists of $n$ variables $X = \set{x_1, \dots, x_n}$ and $m$ clauses $\mathcal{C} = \set{c_1, \dots, c_m}$ 
such that each clause $c_i$ contains at most three literals $\lit{i}{1}, \lit{i}{2}, \lit{i}{3}$. 
Here, $\lit{i}{j} \in \set{x_1, \dots, x_n, \lnot x_1, \dots, \lnot x_n}$ for $1\leq i\leq m$ and $1 \le j \le 3$. 
Given a 3-CNF Boolean formula $\phi$, 
\sat{} asks to decide whether there exists a satisfying assignment for $\phi$. 
In the following, we consider the drawing of the graph corresponding to an instance of \sat{} in the plane. 
The \emph{associated graph} $G(\phi) = (V(\phi), E(\phi))$ of $\phi$ is a graph such that 
$V(\phi) = X \cup \mathcal{C}$, 
$E(\phi) = \inset{\set{x, c}\in X\times \mathcal{C}}{x \in c  {\rm ~or~} \lnot x \in c}$. 
\psat{} is a restriction of \sat{} such that the associated graph of a given formula has a planar embedding; that is, we can draw the graph in the plane in such a way that no edges cross each other. 
It is shown in~\cite{Knuth1992-gu,Lichtenstein1982-bn} that any instance of \psat{} has a \emph{rectilinear representation}, where a rectilinear representation of $\phi$ is a planar drawing of the graph $G(\phi)$ satisfying the following conditions: 
\begin{enumerate}
    \item Vertices in $G(\phi)$ are drawn by rectangles whose sides are parallel to axis, 
    \item rectangles corresponding to variables are drawn on a horizontal line, 
    \item edges in $E(\phi)$ are drawn by vertical segments, and 
    \item edges do not cross each other. 
\end{enumerate}
See \Cref{fig:ex.rectilinear} for an example of a rectilinear representation. 
The decision problem \rlsat{} is defined as follows: Given a 3-CNF formula $\phi$ and its rectilinear representation $D$, 
decide whether there exists a satisfying assignment for $\phi$. 
\rlsat{} is known to be \NP-complete~\cite{Knuth1992-gu}. 

To show the \NP-completeness of \ourproblem{}, 
we construct the corresponding instance $(\mathcal{R}(\phi, D), \succeq(\phi, D), h(\phi, D))$ of \ourproblem{} for an instance $(\phi,D)$ of \rlsat{} 
so that $(\mathcal{R}(\phi, D), \succeq(\phi, D), h(\phi, D))$ is a \Yes-instance of \ourproblem{} if and only if $(\phi, D)$ is a \Yes-instance of \rlsat{}. 
For the construction, we introduce several gadgets. 

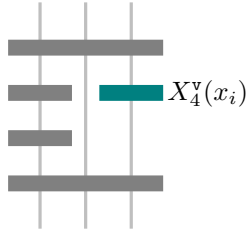
\begin{figure}[h!]
    \centering
    \begin{tikzpicture}[scale=\myscale]
	\foreach \x in {0.5, 1.5, 2.5} {
		\draw [vertical bar] (\x, -1) to (\x, 4);
	}
	\foreach \l / \r / \y in {
		0/3/3,
		0/1/2,
		0/1/1,
		0/3/0
        } {
		\draw [fixed] (\l-0.2, \y) to (\r+0.2, \y); 
	}
	\foreach \l / \r / \y in {
        2/3/2
	} {
		\draw [variable bar] (\l-0.2, \y) to (\r+0.2, \y); 
	}

    \node at (4 + 0.3*\myscale, 2) {$\vgate{4}{x_i}$}; 
\end{tikzpicture}
    \caption{Variable gadgets. In this figure, the corresponding variable is assigned to true. }
    \label{fig:var:gadget}
\end{figure}

\begin{figure}[htbp]
    \centering
    \begin{tikzpicture}[scale=\myscale]
		\foreach \x in {-0.5, 0.5, ..., 8.5} {
				\draw [vertical bar] (\x, -1) to (\x, 8);
			}
		\foreach \l / \r / \y in {
				-1/9/7,
				8/9/6,
				8/9/5,
				-1/2/4, 8/9/4,
				8/9/3,
				8/9/2,
				-1/2/1, 8/9/1,
				-1/9/0
			} {
				\draw [fixed] (\l-0.2, \y) to (\r+0.2, \y);
			}

		\foreach \l / \r / \y in {
				-1/0/6, 
				-1/0/3} {
				\draw [variable bar] (\l-0.2, \y) to (\r+0.2, \y);
			}
		
		\node at (-2 - 0.3*\myscale, 6) {$\bendgate{1}{x_i}$}; 
		\node at (-2 - 0.3*\myscale, 3) {$\bendgate{2}{x_i}$};

		\foreach \l / \r / \y in {
				1/2/6, 
				1/2/3, 
				0/1/5, 
				0/1/2, 
				5/7/6,
				2/5/5, 6/7/5,
				3/4/4, 5/7/4,
				3/5/3, 6/7/3,
				2/4/2, 5/7/2,
				3/5/1 } {
				\draw [inner bar] (\l-0.2, \y) to (\r+0.2, \y);
			}
\end{tikzpicture}
    \caption{Bending gadget. If a green gate in the gadget is on the upper (resp. bottom) level, then the other green gate is also on the upper (resp. bottom) level. This gadget changes the direction of the propagation of the information for a variable assignment. }
    \label{fig:bend:gadget}
\end{figure}
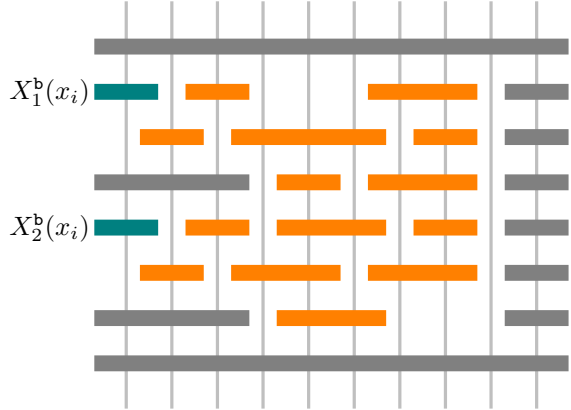

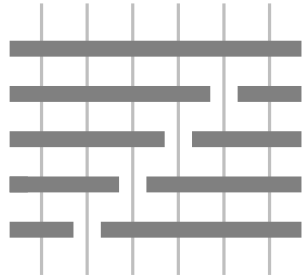
\begin{figure}[h]
    \centering
    \begin{tikzpicture}[scale=\myscale]
	\foreach \x in {0.5, 1.5, ..., 5.5} {
		\draw [vertical bar] (\x, -1) to (\x, 5);
	}
	\foreach \l / \r / \y in {
         0/6/4, 
		0/4/3,
		5/6/3,
		0/3/2,
		4/6/2,
		0/2/1,
		3/6/1,
		0/1/0,
		2/6/0,
        } {
		\draw [fixed] (\l-0.2, \y) to (\r+0.2, \y); 
	}

\end{tikzpicture}
    \caption{Gadget for fixing the ordering of six logical qubits. }
    \label{fig:fix:ordering:gadget}
\end{figure}

Each gadget consists of two types of gates. 
A gate in a gadget is \emph{fixed} if for any packing of the gadget with minimum height, the level of the gate is the same. 
Otherwise, a gate is said to be \emph{movable}. 
Movable gates propagate the choice of the assignment of a variable. 
On the other hand, fixed gates restrict the placement of each movable gate. 
In each gadget, we call the set of fixed gates the \emph{frame} of the gadget. 
In the figures (Figures~\ref{fig:var:gadget}--\ref{fig:fix:ordering:gadget}), the fixed gates are colored gray and the topmost gate is the largest gate. 
In addition, for figures of gadgets other than bending gadgets (Figure~\ref{fig:bend:gadget}), we call green gates on the left side \emph{inputs} and those of the right side \emph{outputs}. 
Intuitively speaking, a gadget receives a variable assignment from the inputs and propagates the information to the outputs. 

\paragraph{(1) Variable gadgets}
This gadget corresponds to a variable in $\phi$.  
The variable gadget for a variable $x_i$ consists of five gates $\vgateset{x_i} = \set{\vgate{1}{x_i}, \dots, \vgate{5}{x_i}}$ defined as follows:  
$\vgate{1}{x_i} =  \vgate{5}{x_i} = \set{1, 2, 3}$,  $\vgate{2}{x_i} =  \vgate{3}{x_i} = \set{1}$, and $\vgate{4}{x_i} =  \set{3}$. 
Moreover, $\vgate{1}{x_i} \prec \vgate{b}{x_i}$ for $b > 1,$ $\vgate{a}{x_i} \prec \vgate{5}{x_i}$ for $a < 5$ and $\vgate{2}{x_i} \prec \vgate{3}{x_i}$. 
See \Cref{fig:var:gadget} for the detail. 
Note that there is no input in variable gadgets. 
Clearly, the height of a minimum packing of $\vgateset{x_i}$ is four. 
In addition, we have only two minimum packings of $\vgateset{x_i}$; we can put $\vgate{4}{x_i}$ at level 2 or 3. 
Note that the level is counted from bottom to top. 
Deciding the level of $\vgate{4}{x_i}$ corresponds to an assignment of the variable $x_i$. 
If $\vgate{4}{x_i}$ is on level 3, then we assign $x_i$ to true. 
On the other hand, if $\vgate{4}{x_i}$ is on level 2, then we assign $x_i$ to false. 

\paragraph{(2) Clause gadgets} 
The clause gadget for a clause $c_y$ consists of 29 gates. 
See \Cref{fig:clause:gadget} for the detail. 
The partial order on those gates are defined to be consistent with this figure.
Note that there is no output in clause gadgets. 
In the gadget, three green gates $X_i$, $X_j$, and $X_k$ correspond to variable assignments. 
For any minimum packing, there are no options for the level of each gray gate. 
On the other hand, we have two options for the level of $X_i$, $X_j$, and $X_k$;   
9 and 10 for $X_i$, 
6 and 7 for $X_j$, and 
2 and 3 for $X_k$.  
If we, respectively, place $X_i$, $X_j$, and $X_k$ on level 9, 6, and 2, 
then there is no packing of the gadget with height 11. 
Otherwise, there is a packing with height 11. 
This implies that at least one literal $x_i$, $x_j$, and $x_k$ must be on the upper level. 
That is, to obtaining a minimum packing with height 11, $c_y$ must be true.

\paragraph{(3) Not gadgets}
To negate a variable, we use a not gadget. 
See \Cref{fig:not:gadget},
which represents gates constituting the gadget and a partial order defined on them.
The height of the minimum packing of this gadget is clearly four. 
If the left green gate is on the upper level of possible choices, that is, assigning true to the corresponding variable, 
then the right green gate is on the lower level, that is, assigning false.

\paragraph{(4) Copying gadgets}
Copying gadgets are used to copy the assignment of a variable $x_i$. 
See \Cref{fig:copy:gadget},
which represents gates constituting the gadget and a partial order defined on them.
As before, we do not have options for the level of each gray gate. 
In addition, clearly, the height of a minimum packing of the gadget is 10. 
For any minimum packing of the gadget, if the green left gate $\copygate{1}{x_i}$ is on the upper level of possible choices, 
then the other green gates $\copygate{2}{x_i}$ and $\copygate{3}{x_i}$ are also on the upper level. 
Similarly, if the left green gate is on the lower level of possible choices, 
then the other green gates are also on the lower level. 
These are the only two packings of the gadgets with height 10. 

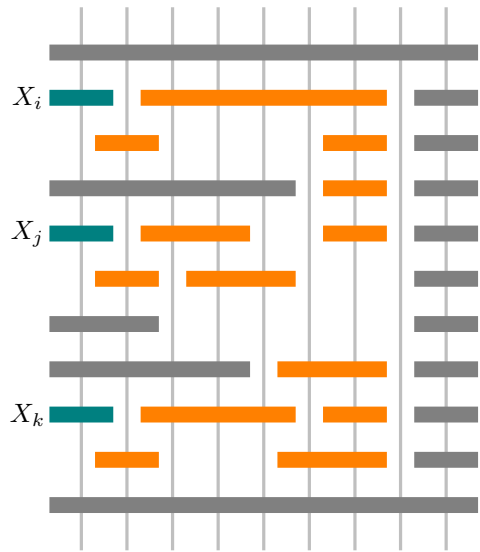
\begin{figure}[h!]
    \centering
    \begin{tikzpicture}[scale=\myscale]
	\foreach \x in {-0.5, 0.5, ..., 7.5} {
		\draw [vertical bar] (\x, -1) to (\x, 11);
	}
	\foreach \l / \r / \y in {
		-1/8/10,
		7/8/9,
		7/8/8,
		-1/4/7, 7/8/7, 
		7/8/6,
		7/8/5,
		-1/1/4, 7/8/4, 
		-1/3/3, 7/8/3, 
		7/8/2,
		7/8/1,
		-1/8/0} {
		\draw [fixed] (\l-0.2, \y) to (\r+0.2, \y); 
	}

	\foreach \l / \r / \y in {
		-1/0/9,
		-1/0/6,
		-1/0/2
	} {
		\draw [variable bar] (\l-0.2, \y) to (\r+0.2, \y); 
	}
	
	\node at (-1.5 - 0.3*\myscale, 9) {$X_i$}; 
	\node at (-1.5 - 0.3*\myscale, 6) {$X_j$}; 
	\node at (-1.5 - 0.3*\myscale, 2) {$X_k$}; 

	\foreach \l / \r / \y in {
		5/6/8,
		5/6/7,
		5/6/6,
		2/4/5,
		4/6/3,
		5/6/2,
		4/6/1,
		1/6/9,
		1/3/6,
		1/4/2, 
		0/1/8,
		0/1/5,
		0/1/1
	} {
		\draw [inner bar] (\l-0.2, \y) to (\r+0.2, \y); 
	}
\end{tikzpicture}
    \caption{An example of a clause gadget. In this example, all variables appearing in the corresponding clause are assigned to true. If all variables are assigned to false, the height of this gadget is at least 12. }
    \label{fig:clause:gadget}
\end{figure}
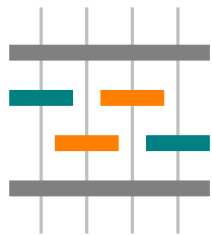
\begin{figure}[h!]
    \centering
    \begin{tikzpicture}[scale=\myscale]
	\foreach \x in {1.5, 2.5, ..., 4.5} {
		\draw [vertical bar] (\x, -1) to (\x, 4);
	}
	\foreach \l / \r / \y in {
		1/5/3,
		1/5/0
        } {
		\draw [fixed] (\l-0.2, \y) to (\r+0.2, \y); 
	}
	\foreach \l / \r / \y in {
        1/2/2, 4/5/1
	} {
		\draw [variable bar] (\l-0.2, \y) to (\r+0.2, \y); 
	}

	\foreach \l / \r / \y in {
        3/4/2, 
        2/3/1 
	} {
		\draw [inner bar] (\l-0.2, \y) to (\r+0.2, \y); 
	}

\end{tikzpicture}
    \caption{Not gadget. The level of the green gates is not the same. }
    \label{fig:not:gadget}
\end{figure}
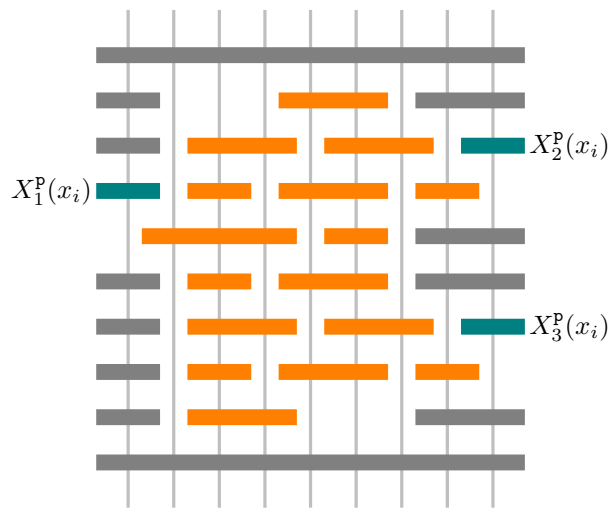
\begin{figure}[h!]
    \centering
    \begin{tikzpicture}[scale=\myscale]
	\foreach \x in {1.5, 2.5, ..., 9.5} {
		\draw [vertical bar] (\x, -1) to (\x, 10);
	}
	\foreach \l / \r / \y in {
        1/10/9,
        1/2/8, 8/10/8,
        1/2/7,
        8/10/5,
        1/2/4, 8/10/4,
        1/2/3,
        1/2/2,
        1/2/1, 8/10/1,
        1/10/0} {
		\draw [fixed] (\l-0.2, \y) to (\r+0.2, \y); 
	}
	\foreach \l / \r / \y in {
        9/10/7,
        1/2/6, 
        9/10/3
	} {
		\draw [variable bar] (\l-0.2, \y) to (\r+0.2, \y); 
	}

    \node at (0  - 0.3*\myscale, 6) {$\copygate{1}{x_i}$}; 
    \node at (11 + 0.3*\myscale, 7) {$\copygate{2}{x_i}$}; 
    \node at (11 + 0.3*\myscale, 3) {$\copygate{3}{x_i}$}; 

	\foreach \l / \r / \y in {
        5/7/8,
        3/5/7, 6/8/7, 
        3/4/6, 5/7/6, 8/9/6,
        2/5/5, 6/7/5,
        3/4/4, 5/7/4,
        3/5/3, 6/8/3, 
        3/4/2, 5/7/2, 8/9/2,
        3/5/1
	} {
		\draw [inner bar] (\l-0.2, \y) to (\r+0.2, \y); 
	}

\end{tikzpicture}
    \caption{Copy gadget. If the left green gate is on the upper (resp. bottom) level, then the right green gates are on the upper (resp. bottom) level. }
    \label{fig:copy:gadget}
\end{figure}
\clearpage
\paragraph{(5) Bending gadgets}
Bending gadgets are used to rotate the direction of propagation of the assignment of a variable $x_i$. 
See~\Cref{fig:bend:gadget},
which represents gates constituting the gadget and a partial order defined on them.
The minimum height is eight. 
As before, for any minimum packing, we have only two options for the level of $\bendgate{1}{x_i}$; 6 or 7. 
If the left green gate $\bendgate{1}{x_i}$ is at level 7, 
then the other green gate $\bendgate{2}{x_i}$ is also at the upper level. 
Similarly,  if the left green gate $\bendgate{1}{x_i}$ is on level 6, 
then the other green gate $\bendgate{2}{x_i}$ is also on the lower level. 
Note that this gadget is utilized to propagate the assignment of a variable to a different level by 
concatenating two bending gadgets so that the new gadget is ``S''-shaped. 

\paragraph{(6) Fixing gadget}
To fix the left-right direction of logical qubits, we add several gates on the bottom of the instance as follows. 
See \Cref{fig:fix:ordering:gadget} for the case that the number of logical qubits is six.  
Gates of this gadget is smaller than any gates in other gadgets. 
Thus, in the packing of the instance, this gadget appears in the bottom. 
For $n$ logical qubits, the minimum height of this gadget is $n-1$. 
For $i= 1, \dots, n-2$, at level $i$, the gadget contains two gates $\fixgate{\ell}{i} = \set{1, \dots, i}, \fixgate{r}{i} = \set{i + 1, \dots, n}$. 
Any other ordering of logical qubits cannot achieve this minimum height. 

\begin{figure*}[htbp]
    \centering
    \begin{tikzpicture}[x=1em, y=1em, scale=\myscale]
\tikzstyle{fixed}=[line width=6*\myscale, color=gray]
\tikzstyle{variable bar}=[line width=6*\myscale, color=teal]
\tikzstyle{inner bar}=[line width=6*\myscale, color=orange]

    \def\vw{0} \def\negvw{1}
    \def\vx{1} \def\negvx{0}
    \def\vy{0} \def\negvy{1}
    \def\vz{1} \def\negvz{0}
    
    \fixgadget{-16}{-28}{59}

    \vgadget{0}{43}{\vw}{w}
    \extendgadget{2}{43}{\vw}{28}

    \vgadget{0}{34}{\vx}{x}

    \copygadget{2}{20}{\vx}{5}{1}
    \extendgadget{11}{35}{\vx}{23}

    \extendgadget{20}{22}{\vx}{14}

    \copygadget{11}{13}{\vx}{2}{1}
    \bendgadget{22}{6}{\vx}{2}{1}
    \extendgadget{-8}{7}{\vx}{28}

    \vgadget{0}{-2}{\vy}{y}
    \copygadget{2}{-8}{\vy}{1}{1}
    \extendgadget{11}{-1}{\vy}{23}

    \bendgadget{13}{-15}{\vy}{2}{1}

    \extendgadget{-8}{-14}{\vy}{19}

    \vgadget{0}{-20}{\vz}{z}
    \extendgadget{2}{-20}{\vx}{28}

    \notgadget{30}{43}{\vw}
    \dummyvgadget{32}{29}{1}
    \cgadget{36}{29}{\negvw}{\vx}{0}{5}{2}{1}{C1}

    \dummyvgadget{-7}{-20}{0}
    \cgadget{-15}{-20}{\vx}{\vy}{0}{18}{2}{0}{C2}

    \notgadget{30}{-20}{\vz}
    \cgadget{36}{-20}{\vx}{\vy}{\negvz}{20}{15}{1}{C3}

    \node at ($(blvw) + (-1.5em, 3)$) {$x_1$};
    \node at ($(blvx) + (-1.5em, 3)$) {$x_2$};
    \node at ($(blvy) + (-1.5em, 3)$) {$x_3$};
    \node at ($(blvz) + (-1.5em, 3)$) {$x_4$};

    \node at ($(urcC1) + (2em, -10)$) {$C_1$};
    \node at ($(blcC2) + (-1.5em, 20)$) {$C_2$};
    \node at ($(urcC3) + (2em, -25)$) {$C_3$};
\end{tikzpicture}
    \caption{The corresponding instance of $\phi_1 =  C_1 \land C_2 \land C_3$ in \Cref{fig:ex.rectilinear}, where $C_1 =(\lnot x_1 \lor x_2), C_2 = (x_2 \lor x_3 ), C_3 = (x_2 \lor x_3 \lor \lnot x_4)$. 
    The figure depicts the case when $(x_1, x_2, x_3, x_4) = (F, T, F, T)$. 
    A blue area surrounded by a dotted line corresponds to a variable, and a yellow are surrounded by a dashed line corresponds to a clause. 
    To propagate the information of the assignment of a variable gadget to a clause gadget, 
    we use copying gadgets and bending gadgets, and extend these gadgets. 
    Additionally, as $C_1$ and $C_2$ contain only two literals, we add dummy variable gadgets to force the third variable to be assigned false.
    $C_1$ and $C_3$ are adjacent to not gadgets since they contain negative literals. 
 }
    \label{fig:reduction:example}
\end{figure*}
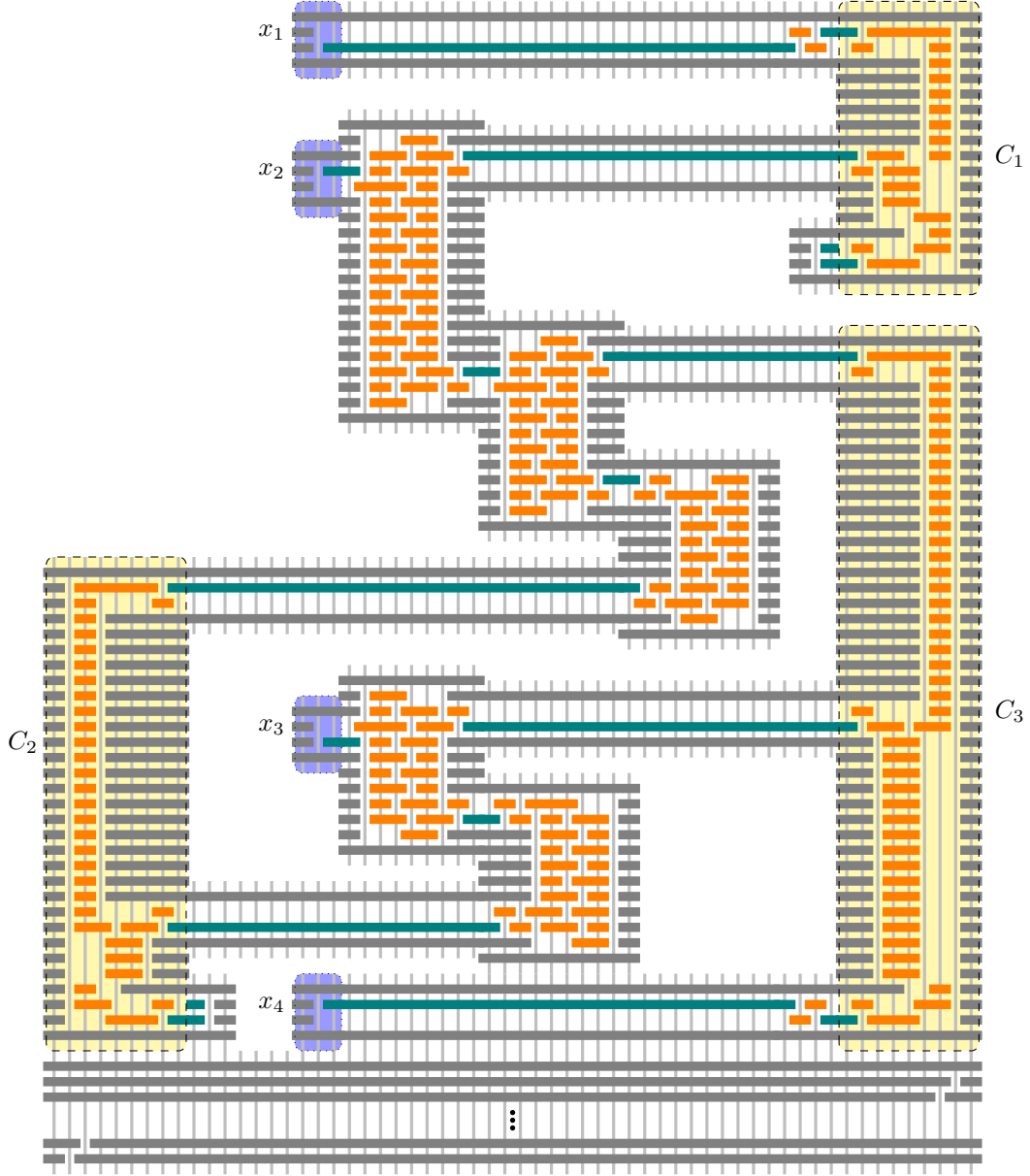

\paragraph{Construction}
Now, we construct the corresponding instance $(\mathcal{R}(\phi),\succeq(\phi), h(\phi))$ for an instance $\phi$ of \rlsat{}. 
We first rotate the rectilinear representation of \rlsat{} by 90 degrees in a clockwise manner. 
Next, we replace each vertex in the representation with the frame of the corresponding gadget. 
Then, we connect the gadgets by using not gadgets, copying gadgets, and bending gadgets and fill the empty place other than the inside of each frame by gray gates.  
Note that, to properly connect these gadgets, we may slightly modify each gadget by inserting gray gates and orange gates, and extending green gates. 
Additionally, if a clause contains less than three literals, then we use a dummy variable gadget to force a unused literal to be assigned false. 
These modifications do not affect the mechanism of each gadget. 
Once the placement of frames are fixed, we can compute the possible minimum height $h(\phi)$ of a packing and the number of vertical lines.
Moreover, we can give a partial order on gates according to the placement of frames. 
\Cref{fig:reduction:example} shows an example of the corresponding instance of $\phi_1$. 
Since the correctness is straightforward from the construction, the following theorem is established. 

\begin{theorem}
$(\mathcal{R}(\phi),\succeq(\phi), h(\phi))$ is a \Yes-instance of \ourproblem{} if and only if $\phi$ is satisfiable. 
\end{theorem}

We should note that we can define a ``2D braiding circuit'' representation and the problem of minimizing the volume of the 2D braiding circuit in a similar way to the 1D braiding circuit representation. The 1D representation is also applicable to quasi 2D arrangements, such as a bi-linear array, in some extent, however the conditions of the margins and overlap for the optimization problem become more complex when one attempts to utilize quantum gates with a more complex shape or to place qubits in a 2D or 3D arrangement. While the complexity of this problem remains an open question, we conjecture that optimization of 2D circuits and higher-dimensional circuits is intractable as it is a generalization of the 1D braiding circuit representation. 
\section{Conclusion}
This paper considers the computational complexity of the problem that must be solved to optimize a quantum circuit by changing the qubit order when using defect braiding, one of the FTQC methods needed for large-scale quantum information processing using the surface code. We proved that this problem is NP-complete by performing a polynomial-time reduction to \rlsat, which corresponds to the equivalent result for lattice surgery\cite{herr2017optimization}. Taken together, these results suggest that optimizing FTQC circuits is costly and that heuristic methods, including approximation algorithms, are essential. As such, the development of such heuristic methods will be important for the realization of FTQC using the surface code, and for the comparison of other error-correction codes and fault-tolerant procedures to determine the most practical and lowest-overhead route to large-scale quantum information processing.

\section*{Acknowledgement}
SN would like to thank Thomas Scruby for useful discussions. This work was supported by JSPS KAKENHI Grant Numbers JP21H04880, JP19K20350, JP22J20882, the MEXT Quantum Leap Flagship Program (MEXT Q-LEAP) Grant Number JPMXS0118069605, and the JST Moonshot R\&D Grant Number JPMJMS2061.

\bibliographystyle{ieeetr}
\bibliography{main}
\end{document}